\documentclass[preprint]{aastex}
\input epsf

\shorttitle{Globular Clusters in NGC 3610}
\shortauthors{Strader et al.}

\begin{document}

\title{Keck Spectroscopy of Globular Clusters in the Elliptical Galaxy NGC 3610}

\author{Jay Strader\altaffilmark{1}, Jean P. Brodie\altaffilmark{1}, Fran\c{c}ois Schweizer\altaffilmark{2}, S{\o}ren S.
Larsen\altaffilmark{1}, and Patrick Seitzer\altaffilmark{3}}
\email{strader@ucolick.org, brodie@ucolick.org, schweizer@ociw.edu, soeren@ucolick.org, pseitzer@umich.edu}

\altaffiltext{1}{UCO/Lick Observatory, University of California, Santa Cruz, CA 95064}
\altaffiltext{2}{Carnegie Observatories, 813 Santa Barbara St., Pasadena, CA 91101}
\altaffiltext{3}{Department of Astronomy, University of Michigan, 818 Dennison Building, Ann Arbor, MI 48109}

\begin{abstract}

We present moderate-resolution Keck spectra of nine candidate globular clusters in the possible merger-remnant elliptical galaxy NGC 3610. Eight of the
objects appear to be bona fide globular clusters of NGC 3610. We find that two of the clusters belong to an old metal-poor population, five to an old
metal-rich population, and only one to an intermediate-age metal-rich population. The estimated age of the intermediate-age cluster is 1--5~Gyr, which is in
agreement with earlier estimates of the merger age, and suggests that this cluster was formed during the merger. However, the presence of five old metal-rich
clusters indicates that a substantial number of the metal-rich clusters in NGC 3610 likely came from the progenitor galaxies, although the global ratio of old
to intermediate-age metal-rich clusters remains very uncertain.

\end{abstract}

\section{Introduction}

NGC 3610 has attracted attention for some time because of its rich fine structure and prominently warped disk, which imply that a significant
dynamical event occurred during the past 5--6~Gyr.  With a much longer time-scale, relaxation would have led to the disappearance of the fine
structure in the inner parts of the galaxy.  Both the enhanced line strength of H$\beta$ \citep{S90} and the bluer-than-average $UBV$ colors
(Schweizer \& Seitzer 1992, hereafter SS92) suggest that NGC 3610 harbors an aging starburst population from several Gyr ago, yet the integrated colors of the
galaxy are not blue enough for it to have undergone a significant burst of star formation during the past 1~Gyr or so \citep{SB90}. NGC 3610 is thus
an excellent candidate for an intermediate-age merger remnant, with an estimated age of $4\pm2.5$ Gyr (SS92; Whitmore et al.~1997, hereafter W97). There is
increasing evidence that major bursts of star formation are typical among gravitationally interacting galaxies, a good recent example being The
Antennae \citep{W99}.

SS92 suggested that elliptical galaxies with intermediate-age stellar populations could be explained by the occurrence of a major disk--disk merger
in their past, which would lend support to the theory that some elliptical galaxies formed in such major mergers (Toomre \& Toomre 1972; Schweizer 1986; 
Ashman \& Zepf 1992; Barnes 1998). The recent slew of evidence for the formation of young massive clusters (YMCs) in many interacting galaxies (see
Whitmore 2002 and Schweizer 2002 for recent reviews) has led naturally to speculations that YMCs are younger versions of the old globular clusters seen
in the Milky Way and other galaxies. If so, a population of intermediate-age clusters could shed light on the suspected merger and might represent
the ``missing link'' between YMCs and present-day old globular clusters (GCs). Few systems of intermediate-age GCs are presently known---probably
the best example being that observed in the merger remnant NGC 1316 (Goudfrooij et al.~2001a, b).  Using a combination of optical/IR
photometric data and spectra of its globular clusters, these authors were able to estimate, with unprecedented precision, a merger age of $3.0\pm0.5$ Gyr.

W97 reported the discovery in NGC 3610 of a population of GCs that was brighter, redder, and more centrally located than the
old, metal-poor population. They speculated that this new subpopulation might represent a set of clusters formed during a major merger
$4\pm2.5$ Gyr ago.  In order to test this speculation, in the present paper we analyze spectra of nine GCs in NGC 3610 and determine
abundances, metallicities, and ages.

\section{Observations and Data Reduction}

The nine GC candidates were observed with the Keck I and II telescopes and the Low-Resolution Imaging Spectrometer \citep{O95} in multislit mode.
Spectra were obtained during two observing runs on 1999 April 7--8 (Keck II) and 2000 February 27--28 (Keck I), respectively.  Seven target objects
for the 1999 run were chosen from the list of GC candidates identified by W97 on images taken with the Wide Field/Planetary Camera~2 on the Hubble
Space Telescope (HST/WFPC2). Two additional objects from that list were added for the 2000 run, based in part upon inspection of new, dithered WFPC2
observations described in \citet[hereafter W02]{W02b}.

During the 1999 run, ten exposures totaling 300 minutes were taken with a single mask over two nights. During the 2000 run, nine exposures totaling
260 minutes were taken with a new mask, also over two nights. All observations were made with a 600 line mm$^{-1}$ grating blazed at 5000 \AA. This
grating provided a reciprocal dispersion of 1.28 \AA\ pixel$^{-1}$ and a spectral resolution of $\sim$6 \AA. The total wavelength range covered by
the CCD detector was generally $3600-6000$ \AA, but the usable range, especially at the blue end of the spectrum, was often shorter by several
hundred Angstroms due to the differences in the location of individual slitlets within the mask.

The standard data reduction was performed using IRAF\footnote{IRAF is distributed by the National Optical Astronomy Observatories, which are
operated by the Association of Universities for Research in Astronomy, Inc., under cooperative agreement with the National Science Foundation.} and
the data analysis package BAOLAB, written by one of us (SL). Raw images were debiased and then flatfielded, using a normalized composite sky flat.
After cosmic-ray removal using the IRAF task \emph{cosmicrays}, the BAOLAB tasks \emph{xdistmap} and \emph{xdistapp} were used to rectify the
science spectra, since several of the slitlets were slightly tilted with respect to the spatial axis. This tilting was done to allow more cluster
candidates to fit on the slitmask. The extracted spectra were then wavelength calibrated using spectra of comparison arc lamps.

After this wavelength calibration, the science spectra were shifted by small amounts in wavelength in order to bring several bright sky lines,
principally those of oxygen, into agreement with their laboratory wavelengths.  The spectra were then coadded with sigma clipping. The coadded
spectra were flux-calibrated using observations of the flux standard PG 0823+546, chosen from \citet{M88}. To assess the quality of the
observations, the signal-to-noise ratio (S/N) of the spectra was measured in the range $4700-5300$ \AA, where most features of interest lie.

Table 1 lists basic data for the nine clusters in our sample.  The positional and photometric data are taken from W02, while the heliocentric
radial velocities are based on our own measurements described in \S 3. No extinction corrections were applied since \citet{BH84} give $A_{V}$ =
0.00 mag and \citet{S98} give $A_{V}$ = 0.03 mag.

One item of note is that during the 2000 run, the important H$\beta$ feature in the spectrum of object W3 was marred by a block of bad pixels.
Since no effective correction could be applied, our subsequent analysis of W3 includes data solely from the 1999 run. This problem is not overly
disconcerting, however, since the quality of the 1999 spectrum alone is relatively high (S/N $\sim 25$).

Spectra of a representative sample of clusters (W6, W9, and W10) are presented in Figure 1.

\section{Radial Velocities}

\subsection{Velocity Measurements}

The radial velocities of individual clusters were measured by cross-correlating the cluster spectra with spectra of the radial velocity standards HR
1805 and HR 3905 using the IRAF task \emph{fxcor}.

For cluster W26* no clear cross-correlation peak with the standard stars could be found. While a reasonable value was obtained by cross-correlating
the cluster spectrum with spectra of several other clusters, the correlation coefficient of the match was low, indicating the measured radial
velocity may have been spurious. Though these difficulties could be due to the relatively low S/N of the cluster, comparison of the smoothed
spectrum of W26* with those of W6 and W10 revealed little resemblance. The spectrum of W26* shows none of the strong lines (e.g. H$\beta$, Mg)
present in the other clusters, nor does it show an H+K line break. Furthermore, the strength of its spectrum increases towards the UV, unlike those
of the other clusters. Visual examination of the WFPC2 images show that the cluster appears superimposed on an extended red object, which may be a
background galaxy. If indeed the cluster is real, the observed spectrum may be dominated by that of the galaxy. For these reasons, we have
excluded W26* from further analysis, and it does not appear in any of the subsequent tables or figures.

The mean heliocentric radial velocity of the eight clusters is $1769\pm15$ km s$^{-1}$, with a velocity dispersion of $45\pm25$ km s$^{-1}$. This
mean velocity is in good agreement with the value $1787\pm48$ km s$^{-1}$ for the systemic velocity of NGC 3610 from the Third Reference Catalogue of Bright
Galaxies \citep{RC3}, but does differ significantly from the more recent value of $1696\pm17$ km s$^{-1}$ listed in the Updated Zwicky Catalogue
\citep{F99}.

There is no clear radial trend in the eight cluster velocities, nor do there appear to be trends with $V$ magnitude, color, or (based upon data in
\S 4) [Fe/H]. There are no distinct kinematic subgroups visible either. With such a small sample size, these null results are not unexpected.

\subsection{Galactic Mass}

The estimated errors of the radial velocities are small enough that we can make a rough estimate of the mass of NGC 3610 lying within the orbits of
our cluster sample. We use the new tracer mass estimator (Evans et al.~2002), which has numerous advantages over the commonly-used projected mass
estimator (Bahcall \& Tremaine 1981; Heisler et al.~1985). These include: no assumption that the tracers follow the dark matter distribution, the use of an
inner and outer radius for the tracer population, and a smaller dependence on the exact orbital shapes. (Evans et al.~find that assuming isotropic orbits
only introduces a 30\% uncertainty in the final mass estimate). If we assume that the potential is isothermal and the GC population is isotropic and falls
off as $\sim r^{-4}$, then the equation for the tracer mass estimator is:

\begin{equation}
        M = \frac{16}{\pi GN}\frac{(r_{\rm{in}}/r_{\rm{out}})^{-1}-1}{\rm{log}(\it{r}_{\rm{out}}/\it{r}_{\rm{in}})}\sum_{i}^N r_{i}(V_{i} - \bar V)^{2}  
\end{equation}
\noindent
where $r_{\rm{in}}$ and $r_{\rm{out}}$ are the inner and outer radii, respectively, of our tracer population. We find that $M = 8.6 \times 10^{10}
\,\textrm{M}_{\odot}$ within a radius of 13 kpc. The sources of error in this estimate are numerous, including measuring errors in the radial velocities, the
limited sample size, our lack of knowledge of the true GC orbits and radial distribution, and the assumption of an isothermal potential (especially in the
context of the inner regions of a merger remnant). The uncertainties are great enough that it makes little sense to give a formal error estimate, whence we
will express our mass estimate as $\sim 8 \times 10^{10} \,\textrm{M}_{\odot}$, with the caveat that this value is uncertain by at least a factor of two.

\section{Metallicities, Abundances, and Ages}

\subsection{Metallicities}

\citet[hereafter BH90]{BH90} defined a procedure for measuring the metallicity of globular clusters by taking a weighted mean of various elemental
absorption-line indices sensitive to [Fe/H]. Indices defined over a wavelength range of interest are calculated with respect to a pseudocontinuum,
defined using regions to either side of the feature passband.

Before measurement of any features, the spectra were flux calibrated as described in \S 2. The minor role of flux calibration in measuring indices
has been documented extensively \citep{F85, KP98, LB02}. Errors due to flux calibration are extremely small, amounting to $\sim$5\% of the smallest
error bars in our study. No resolution corrections were made for measuring BH90 metallicity indices, since the BH90 calibration primarily used data
near the resolution of our data or better. The spectra were then corrected for radial-velocity shifts using the values given in Table 1.

We measured [Fe/H] for our eight clusters according to the formulae in BH90, with the exception of a slight modification in calculating the
variance, as described in \citet{LB02}. The resulting individual index [Fe/H] estimates are listed in Table 2.

To provide a check on these spectroscopic metallicity estimates, we also calculated photometric [Fe/H] values according to
the \citet{KP98} equation:
\begin{equation}
	\textrm{[Fe/H]} = (-4.50\pm0.30) + (3.27\pm0.32)(V-I)
\end{equation}

The photometric and spectroscopic [Fe/H] values are given in Table 3, and the estimates compared in Figure 2. Seven of the eight clusters fall near a 1:1
relation, indicating good agreement between the two value. The one outlier is W30, which has a low-S/N spectrum. The supersolar spectroscopic [Fe/H] value for
this cluster is particularly suspect because of the wide spread in the individual metallicity indices, which range from $\textrm{[Fe/H]}_{\Delta} = -2.00$ to
$\textrm{[Fe/H]}_{CNB}$ = 3.66. This suggests that while it may be possible that the cluster has a metallicity near solar (such clusters have recently been
discovered in NGC 1399 by Forbes et al.~2001), the large error bars make it impossible to put any strong constraints on its metallicity at this time.

Setting aside the cluster W30 due to its large error bars, and using a dividing line of [Fe/H] = $-$1.00 between metal-rich and metal-poor clusters,
the mean metallicities of these two subpopulations are [Fe/H] = $-0.69\pm0.11$ and $-1.28\pm0.15$, respectively. These values are certainly
consistent with the ``standard'' values of $-$0.5 and $-$1.5 \citep{L01, KW01}, especially considering the small-number statistics operating. There
is generally a great deal of variation in the location of the metal-rich peak among galaxies, and \citet{L01} find that the scatter in the color of
the metal-rich peak is 0.04 mag, equivalent to $\sim$0.13 dex in metallicity, using the \citet{KP98} relation.

\subsection{Lick/IDS Indices}

Lick/IDS indices have evolved from the original work of \citet[hereafter B84]{B84}, with several authors providing refinements to the system,
usually in the form of additional indices that cover new features or define the pseudocontinuum regions of an existing feature passband differently
\citep{W94, T98}. These refinements have been accompanied by new sets of cluster-evolution models that predict how various Lick/IDS indices (e.g.,
H$\beta$, Mg$_{2}$, Fe5335) evolve in age--metallicity space \citep{BC02, MT00, M01, W94, AB95}. The refined spectroscopic indices and new models
are especially important in the study of intermediate-age clusters, since the colors of a 3--4 Gyr old population can easily blend with those of a
much older metal-rich component, rendering photometric data insufficient to determine ages (Worthey et al.~1994; W97). In passing we note that
this degeneracy can be partially lifted through the use of infrared photometry. For example, Puzia et al. (2002) demonstrate that subpopulations
which separate cleanly in a $VIK$ color-color diagram would blend together in the common $VI$ color-magnitude diagram.

For the measurement of Lick/IDS indices we had to correct for spectral resolution.  Since the calibration data came from sources with a spectral
resolution range of 5--12 \AA, we convolved our spectra with a 5-pixel (equivalent to 6.4 \AA) Gaussian kernel as a ``middle-of-the-road'' approach
to more closely match the resolution of the original Lick data. Although some previous studies (e.g., Kissler-Patig et al.~1998) have found
little or no resolution degradation necessary when compared to the standard values in \citet{W94}, other studies (e.g., Larsen \& Brodie 2002) have
found offsets of $0.2-0.3$ \AA\ with respect to those standard values when measuring indices in unsmoothed LRIS spectra. These differences are
comparable to the error bars on many of our index measurements. Thus, we decided that smoothing our spectra was the best course of action. It is,
however, worthwhile to recall the \citet{KP98} warning that ``care should be taken in the intercomparison of measurements of various groups.''

For this study, we measured a set of Lick/IDS indices for comparison to theoretical models. These included the higher-order Balmer indices H$\gamma_{A}$
and H$\delta_{A}$ (Worthey \& Ottaviani 1997). Table 4 gives these indices for all eight clusters.

\subsection{Ages}

To estimate the ages of the GCs in our sample, we plotted various combinations of the age-sensitive indices (e.g., H$\beta$ and H$\gamma$) against
the metallicity features (e.g., Mg$_{2}$, Mgb, Fe5270, and Fe5335). As an example, Figure 3 shows H$\beta$ plotted against the averaged Fe5270$+$Fe5335
index. Superimposed on this plot are Maraston theoretical isochrones and isometallicity lines.

Figure 3 clearly shows three distinct subgroups of clusters: an old metal-poor population, an old metal-rich population, and a single metal-rich
cluster (W6) of intermediate age. All of the old clusters fall below the 14 Gyr isochrone, but the model age uncertainties are great enough that
this is not particularly worrisome. Indeed, a general feature of simple stellar population (SSP) models is that while the absolute ages are
uncertain, relative ages are usually consistent. Within the error bars, all seven old clusters seem to be coeval.

While the Maraston (and other SSP) models are intended mainly to represent old stellar populations, and thus may be less accurate for intermediate-age
clusters, it is clear that the cluster W6 cannot be older than $\sim$5 Gyr. The absence of strong Balmer lines implies that the cluster also cannot be
$\la\,$1~Gyr old.  Therefore, the age of W6 seems reasonably constrained to lie in the range 1--5 Gyr. A literal interpretation of Figure 3 suggests an age of
$2.5^{+1.5}_{-0.9}$ Gyr.

These arguments are confirmed by an H$\beta$ vs.\ Mg$_2$ diagram (Fig.~4) and an H$\beta$ vs. [MgFe] diagram (Fig.~5). Though the old
metal-poor and metal-rich subpopulations appear blurred, the general result is the same: one of the clusters is relatively young, while the rest are old
and---within the error bars---coeval. (No confirmation could be made using the age-sensitive indices H$\gamma_{A}$ and H$\delta_{A}$, as their error bars are
too large.)  One advantage of using the combined [MgFe] index is that it is insensitive to variations in [$\alpha$/Fe].

The absolute magnitude of cluster W6 is consistent with it being relatively young; at $M_{V} = -10.60$ (W02), it is the sixth-brightest cluster in NGC 3610.  
This is more luminous than any Milky Way GC, and nearly as luminous as the brightest GC in M31, G1, which has M$_{V}$ = $-$10.85 \citep{R94} (However,
\citet{BPB02} have recently presented evidence that the heavily reddened cluster 037-B327 may have M$_{V}$ = $-$11.74). Excepting the outlying [Fe/H] value of
cluster W30, W6 is also the most metal-rich GC in our sample with [Fe/H] = $-0.29\pm0.19$. Note, however, that the BH90 calibration is based on a sample of
\emph{old} galactic and M31 GCs, whence this metallicity estimate could be significantly less accurate than the formal error bars indicate. Lick/IDS Fe and Mg
indices for W6, interpreted with Maraston models (Figs. 3 \& 4), indicate the metallicity of the cluster may be solar or supersolar, much higher than the
previous estimate.

Our age estimate for cluster W6 is in good agreement with the merger age of $4\pm2.5$ Gyr deduced by SS92 and W97, and provides evidence in favor of
the idea that W6 is an intermediate-age object formed in a merger several Gyr ago.

Figure 6 shows a color-magnitude diagram of all GCs detected in the W02 survey, with our eight clusters (and the intermediate-age cluster W6) specially
marked.

\subsection{[$\alpha$/Fe] Abundances}

To estimate [$\alpha$/Fe] for the clusters in our sample, we used the new $\alpha$-enhanced models of Thomas et al. (2002). Figure 7 is an Mgb vs. $<$Fe$>$
diagram for the seven old GCs, overplotted with 14 Gyr model isochrones for [$\alpha$/Fe] = 0.0 to +0.5. Though there is substantial scatter among the points,
there seems to be a hint of a spread in [$\alpha$/Fe], ranging from $\sim 0.0$ to +0.3. The solar $\alpha$-element abundance ratios for several old clusters
suggest that either the GCs formed over a period of several Gyr, during which time there was substantial enrichment of the ISM by Type Ia supernovae
(SNe), or the GCs with differing $\alpha$-enhancement came from different progenitor galaxies; within each individual galaxy, GC formation took place on a
short timescale. The latter scenario offers the intriguing possibility of associating individual GCs with their parent galaxies.

Figure 8 shows a similar diagram for the intermediate-age GC W6, with 3 Gyr isochrones overplotted. Interestingly, W6 seems to have [$\alpha$/Fe] $\sim +0.3$.
A merger-induced starburst, created as gas from each of the progenitor disks is funneled to the center of the potential well, would be expected to enrich the
ISM with $\alpha$-elements through Type II SNe within a few hundred Myr. Assuming that the gas started with [$\alpha$/Fe] $\sim +0.0$, this enrichment would
have been substantial, raising the \emph{absolute} $\alpha$-element abundance of the ISM from half-solar to solar. If so, W6 must then have formed before the
Type Ia SNe had time to bring [$\alpha$/Fe] back up to solar, suggesting that the cluster was formed in the early days of the merger.

Alternatively, if W6 formed with a very top-heavy IMF, internal enrichment could be responsible for the supersolar [$\alpha$/Fe] ratio. Whether such GCs exist
in other galaxies is unclear; Brodie et al.~(1998) and Larsen \& Brodie (2002) suggest that the anomalously strong Balmer lines in a few young GCs in NGC 1275
and NGC 1023, respectively, could be caused by a top-heavy IMF. However, the significance of these results is colored by the uncertainties in SSP models at
such young ages.

\section{The Nature of the Merger}

Even before GCs were studied in NGC 3610, several pieces of evidence pointed toward this E5 galaxy being the remnant of a major merger.  Both
its unusually rich fine structure \citep{SS90} and warped central disk (W97) strongly suggest that it is dynamically young.  Given the
estimated merger age of $\sim$4 Gyr, the presence of sharp-edged ripples (``shells''), luminous plumes, and an exceptionally boxy halo
(Schweizer 1998, esp.\ Fig.~41) can hardly be attributed to the infall of a minor companion many orbital periods ago. Rather, such a halo shape
and dynamically cold features tend to be signatures of merged disk galaxies and of the delayed return of their tidally ejected material (see Barnes 1998 for a 
review).

Supporting the morphological evidence in this otherwise normal small-group elliptical is strong evidence for an aging starburst: spectral
indices indicate enhanced H$\beta$ absorption and weakened Mg and CN features, as expected from an aging starburst superposed on old stellar
populations \citep{S90}. From the bluer-than-average $UBV$ colors and a heuristic merger model, SS92 estimate a likely merger age of around 4
Gyr.  Near-infrared $JHK$ photometry and HST/STIS spectra in the space UV indicate the presence of a significant intermediate-age population as
well (Silva \& Bothun 1998; Spinrad 1997). Contrary to the claims by Silva \& Bothun, the presence of a central stellar disk is perfectly compatible
with a major disk--disk merger origin of this galaxy: gas-rich mergers tend to form first gaseous, and then stellar, central disks from tidally
ejected and returning gas (Hernquist \& Barnes 1991; Barnes 2002), as directly observed in some nearby remnants of recent mergers.

The study of GCs in NGC 3610 adds interesting new information plus some question marks to this proposed disk-merger formation scenario.

Like the majority of elliptical galaxies, NGC 3610 harbors a GC system with a bimodal color distribution indicative of at least two cluster
subpopulations (W97; W02).  The spatial distribution of the red subpopulation appears more centrally concentrated than that of the blue
subpopulation, suggesting that additional gaseous dissipation occurred between the formation of the metal-poor and metal-rich clusters.  What is
unusual about the GC system of NGC 3610 is that the brightest red GCs are about 0.7~mag more luminous in $V$ than the brightest blue clusters (though
spectroscopy is necessary for certain discrimination of metal-rich and metal-poor clusters). This magnitude difference, together with the measured color
difference $\Delta(V-I)$ between the blue and red peaks of the color distribution, suggests that at least some of the red GCs may be only about $4\pm 2.5$
Gyr old (see Figs.~13 \& 15 in W97).

The luminosity functions of the two GC subpopulations appear to support a significant mean-age difference: while the luminosity function of the
blue GCs is approximately lognormal, that of the red GCs is well represented by a power law with index $\alpha \approx -1.8$ (W02).  This
power-law shape suggests that the red subpopulation is dynamically much less evolved and, hence, significantly younger than the blue
subpopulation. (However, see below for a caveat on the red LF).

Our spectroscopic observations now add more detailed age and metallicity information, though only for eight GCs, all lying beyond a projected
radius of 3 kpc from the center.

First, one of the two sampled GCs nearest the center appears indeed to be of intermediate age, with an estimated age of 1--5 Gyr. This strengthens the cases
previously made for the presence of stellar and GC populations of intermediate age in NGC 3610.  According to the BH90 calibration (Table 3) and the Lick/IDS
indices interpreted with Maraston models (Figs.~3 \& 4) the metallicity of this one cluster is [Fe/H] $\ga\,-$0.3 (see \S 4.3), while the Mg abundance may be
solar or slightly higher (Fig.~4). The metallicity and Mg abundance argue strongly against this cluster having formed from the metal-poor gas of an infalling
dwarf galaxy. For example, the mean metallicity of 14 LMC clusters in the age range 1.5--4 Gyr is only $\langle$[Fe/H]$\rangle = -0.74$, with a scatter of
$\sigma = 0.22$ (Leonardi 2000). Hence, any LMC-like dwarf falling into NGC 3610 1.5--4 Gyr ago would seem unlikely to contribute or produce clusters of the
metallicity observed in W6. Rather, a metallicity of [Fe/H]$ \ga -0.3$ is in accord with the gaseous metallicities that one might expect in some average Sb
or Sc spirals about 2--4 Gyr ago (e.g., Pagel \& Tautvai\v{s}ien\.{e} 1995).

Second, both Figs.~3 and 4 suggest that there is a significant age difference between the one intermediate-age cluster W6 and the other five
``metal-rich'' clusters, the latter apparently being older than $\sim$10 Gyr.  In accord with this age difference, the metallicities of these
five clusters are significantly lower as well, averaging perhaps around [Fe/H] $\approx -0.7$ to $-$1.0 (see also Table 3). Thus, the red
subpopulation of GCs in NGC 3610 seems to contain a broad range of cluster ages and metallicities. Specifically, the ages of the six red clusters suggest
that there were at least two major formation episodes of metal-enriched GCs. This raises a currently unanswerable question: What fraction of the red GCs
formed during the suspected major-merger event $\sim$4 Gyr ago? This is an important factor in determining how the specific GC frequency in this galaxy will
evolve over a Hubble time.

With our present sample of only six red clusters, of which one appears to be of intermediate age, the relative numbers of old and intermediate-age
objects are poorly constrained. Furthermore, the spectroscopic sample is clearly biased toward the outer halo, while the intermediate-age GCs
may be preferentially located in the inner parts of the galaxy. 
 
While the power-law luminosity function of the red GCs would seem to indicate that many of these clusters may be of intermediate age rather
than old, one would clearly expect a number of metal-rich, old clusters from the progenitor galaxies. In the Local Group (dominated by the GC
systems of the Milky Way and M31), the overall ratio of metal-poor to metal-rich GCs is about 2.5:1 \citep{F00}. Thus, we might expect at least
30\% of the old GCs in NGC 3610 to be metal-rich, although the number of metal-rich clusters probably depends on the Hubble type of the
progenitor spirals. Comparison of the relative numbers of old and intermediate-age objects is further complicated by their different luminosity
functions, age fading, and the fact that a portion of the intermediate-age objects are going to disintegrate as their LF evolves towards a more nearly
log-normal shape \citep{FZ01, V01}. Spectroscopy of a larger sample of GCs extending to smaller radii, where---unfortunately---the galaxy
background becomes a major obstacle, will be necessary to further constrain the relative numbers of old and intermediate-age objects.

\section{Conclusions}

We have observed nine globular cluster candidates in the disturbed elliptical galaxy NGC 3610, which is thought to be a merger remnant.  Eight of the nine
objects appear to be bona fide globular clusters belonging to NGC 3610. Seven of these clusters fall into the common old metal-poor and old metal-rich
subpopulations. However, there is considerable evidence for an intermediate-age metal-rich cluster that likely formed in the merger event 1--5 Gyr ago. From
the cluster kinematics and using the tracer mass estimator, we estimate the mass of the galaxy within a radius of 13 kpc to be $\sim 8 \times 10^{10}
\,\textrm{M}_{\odot}$. Finally, we point out that the ages and metallicities of clusters in the red subpopulation suggest at least two cluster formation
episodes, although the relative importance of each of these two formation epochs in creating the current metal-rich cluster population is unknown.

\acknowledgements

We thank Linda Schroder for help with the observations.  We also gratefully acknowledge support by the National Science Foundation
through Grants AST-9900732, AST-0206139 to JPB and AST-9900742, AST-0205994 to FS. We acknowledge the helpful comments of the anonymous referee.

\newpage

\epsfxsize=14cm   
\epsfbox{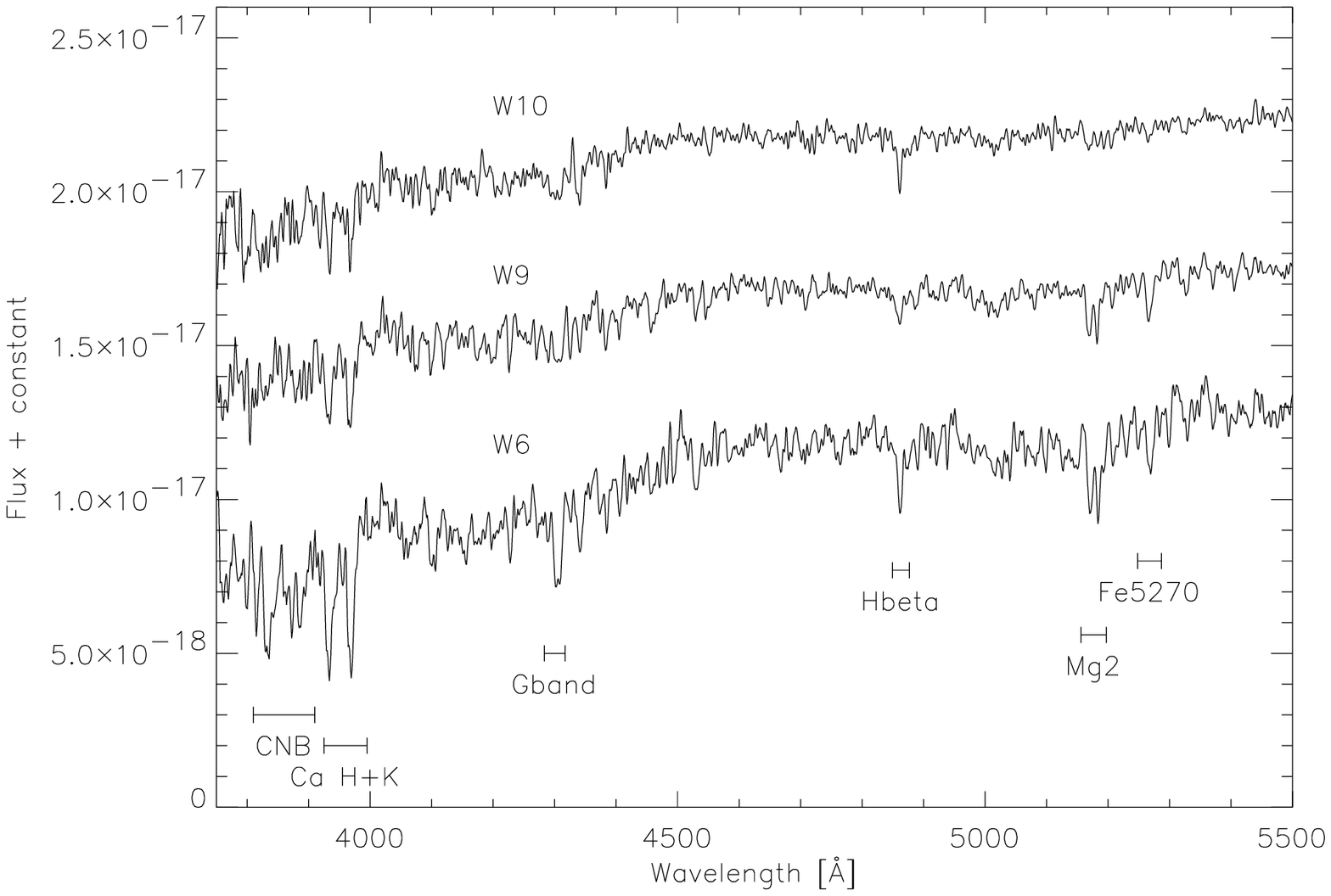}
\figcaption[strader.fig1.ps]{\label{fig:spec}Spectra of clusters W6 (intermediate-age metal-rich), W9 (old metal-rich), and W10
(old metal-poor). Notice the strong H$\beta$ line in W6 and the decreasing strength of metal lines from W6--W10.}

\newpage

\epsfxsize=14cm
\epsfbox{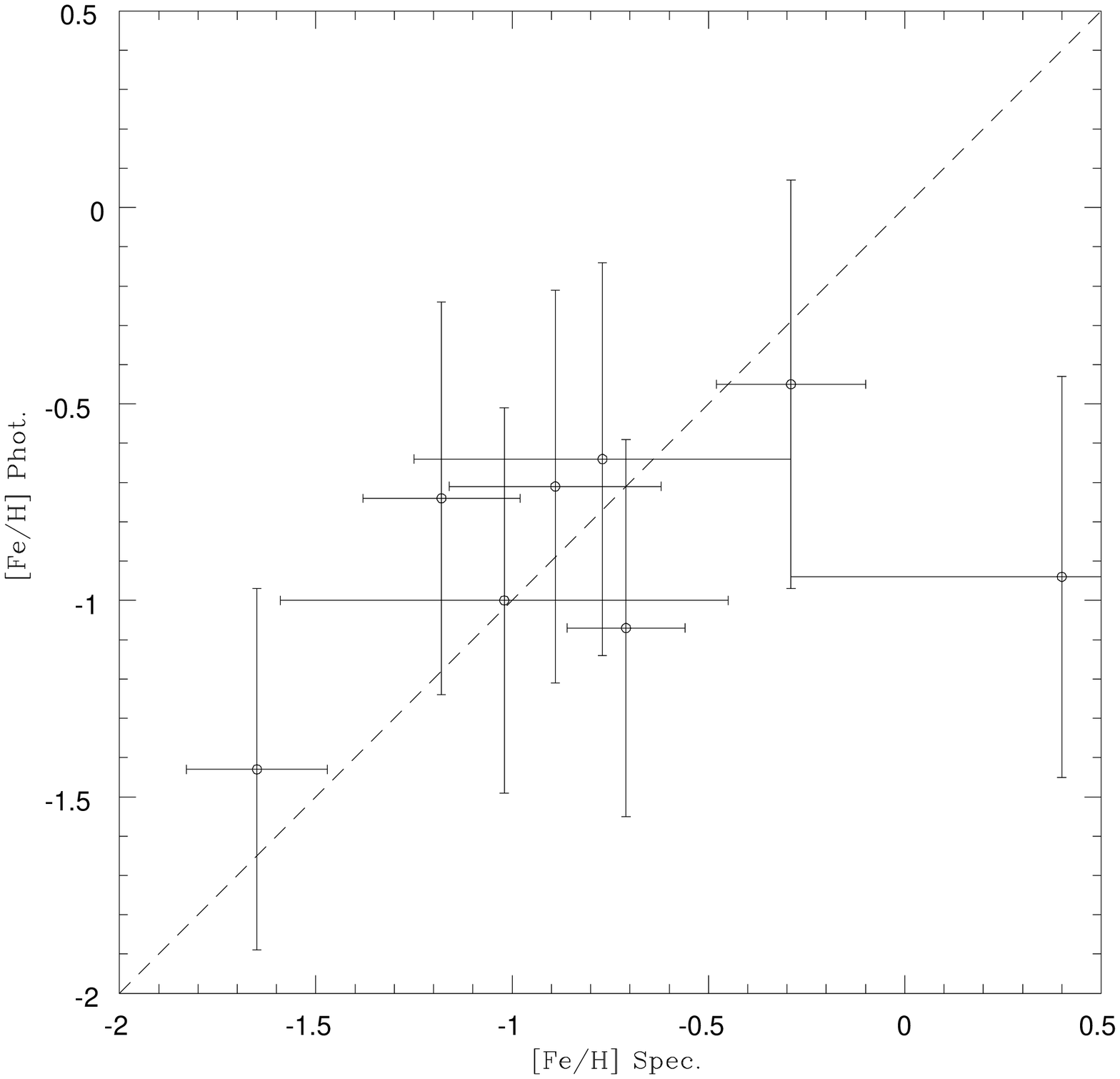}
\figcaption[strader.fig2.ps]{\label{fig:metal}Comparison of spectroscopic and photometric [Fe/H] estimates.}

\newpage

\epsfxsize=14cm
\epsfbox{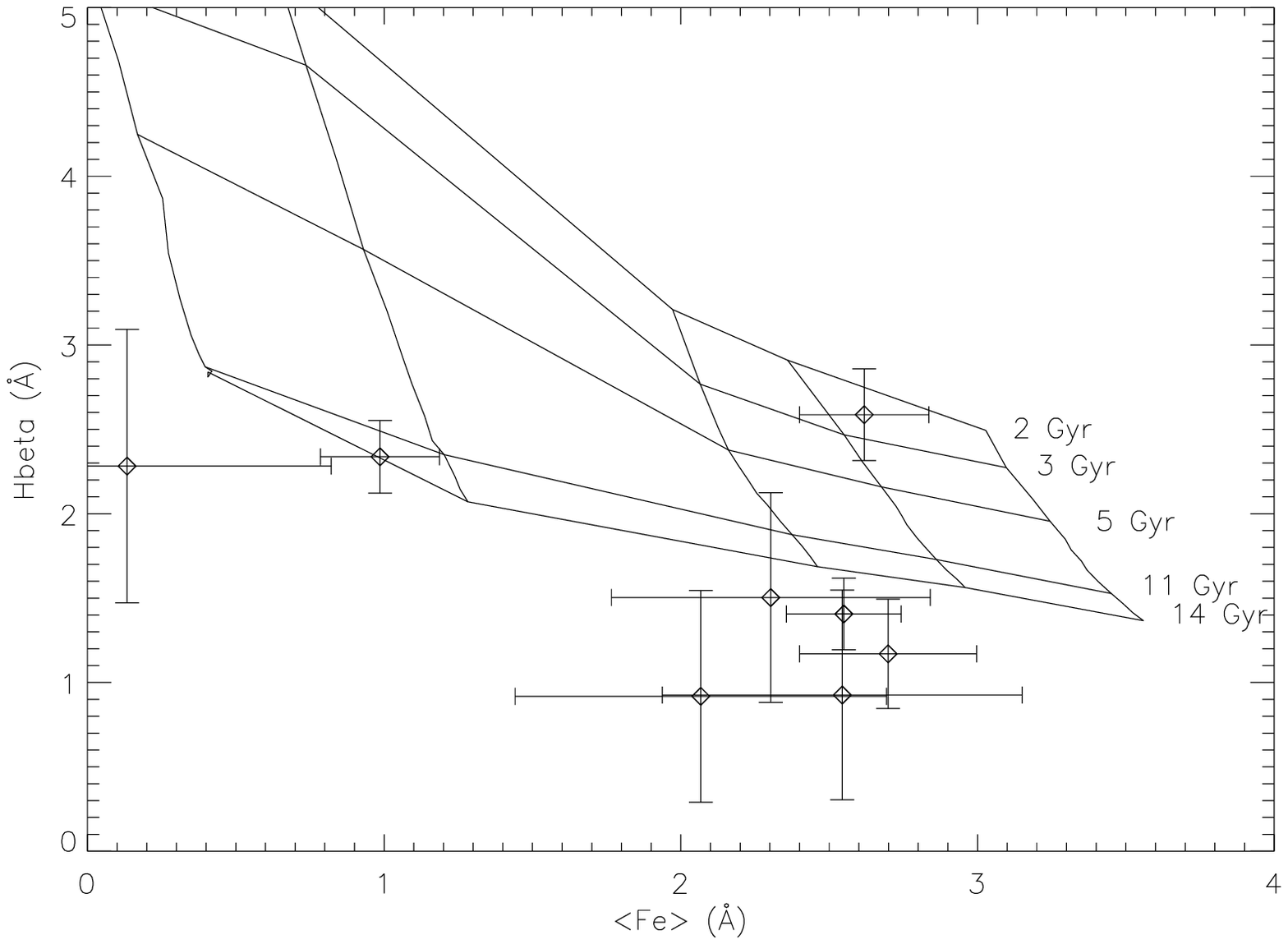}
\figcaption[strader.fig3.ps]{\label{fig:fe_hb}Plot of H$\beta$ vs.\ a composite Fe index, with a grid of model isochrones and
isometallicity lines by Maraston superposed. From left to right, the isometallicity lines represent [Fe/H] = $-$2.25, $-$1.35, $-$0.33, 0.00, 0.35.}

\newpage

\epsfxsize=14cm
\epsfbox{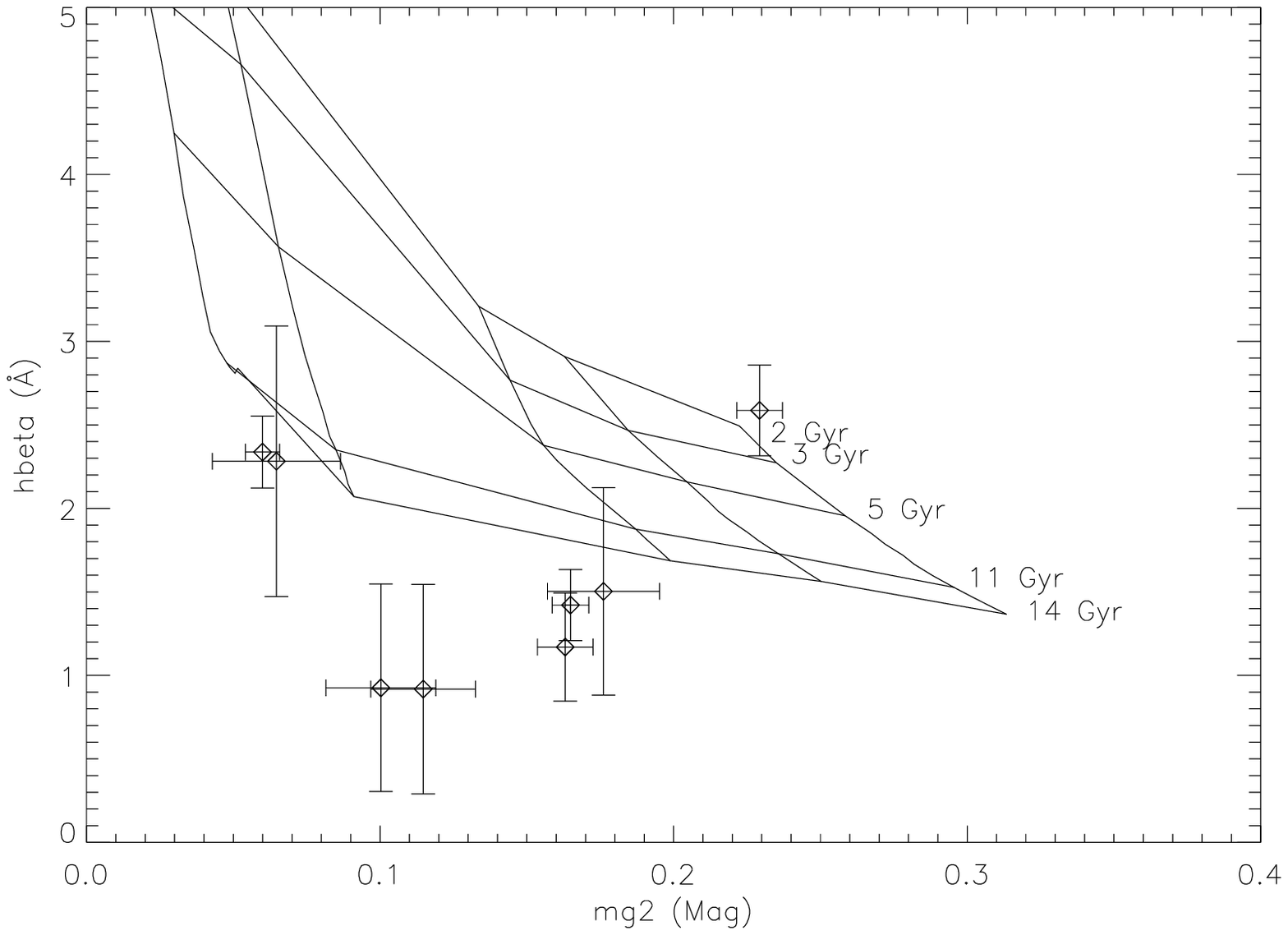}
\figcaption[strader.fig4.ps]{\label{fig:mg2_hb}Plot of H$\beta$ vs.\ Mg$_{2}$, with a grid of model isochrones and
isometallicity lines by Maraston superposed.  From left to right, the isometallicity lines represent [Fe/H] = $-$2.25, $-$1.35, $-$0.33, 0.00, 0.35.}

\newpage

\epsfxsize=14cm
\epsfbox{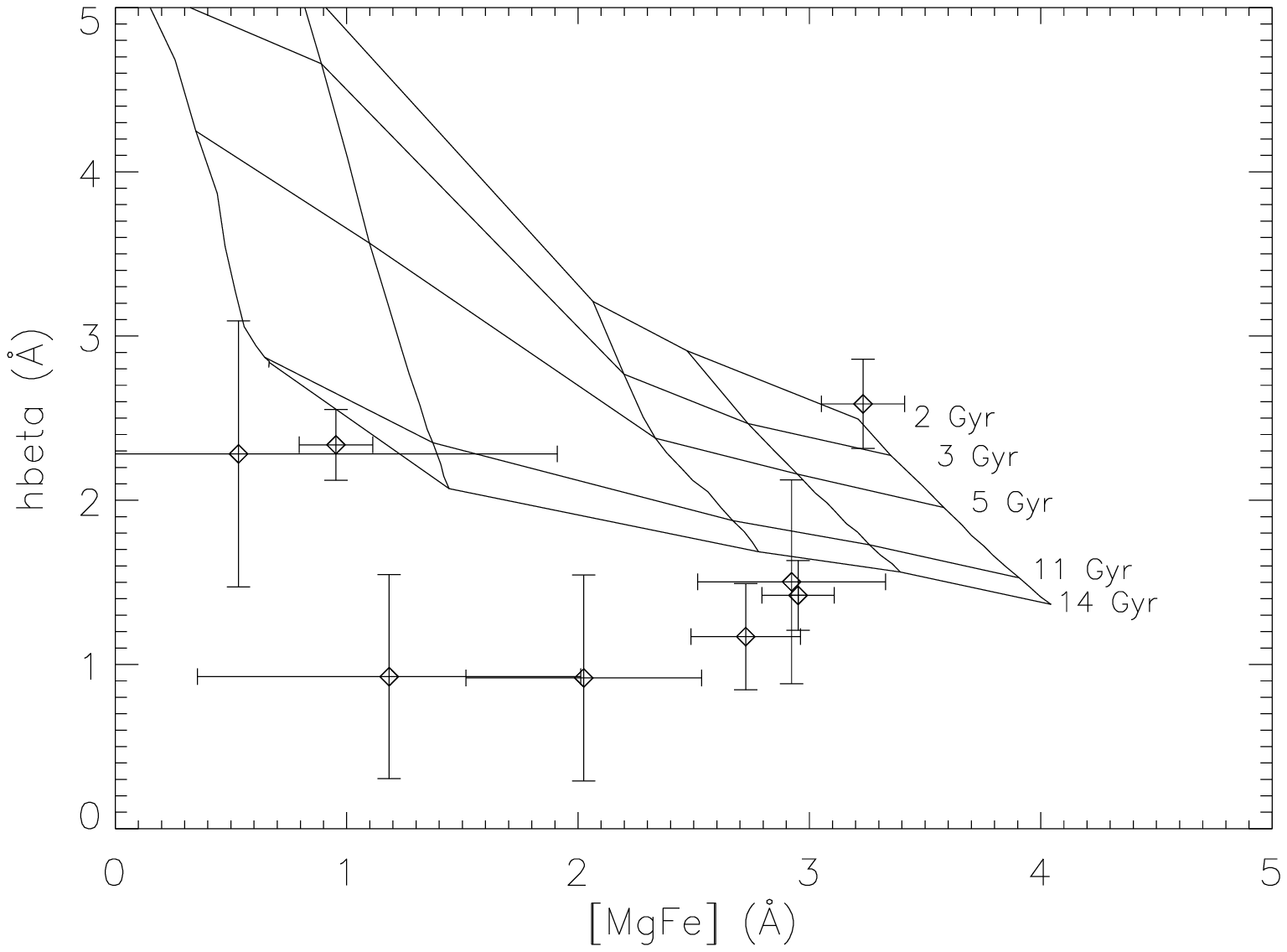}
\figcaption[strader.fig5.ps]{\label{fig:mgfe_hb2}Plot of H$\beta$ vs. [MgFe], with a grid of model isochrones and isometallicity lines by Maraston
superposed. [MgFe] is a  composite index insensitive to [$\alpha$/Fe] variations. From left to right, the isometallicity lines represent [Fe/H] = $-$2.25,
$-$1.35, $-$0.33, 0.00, 0.35.}

\epsfxsize=16cm
\epsfbox{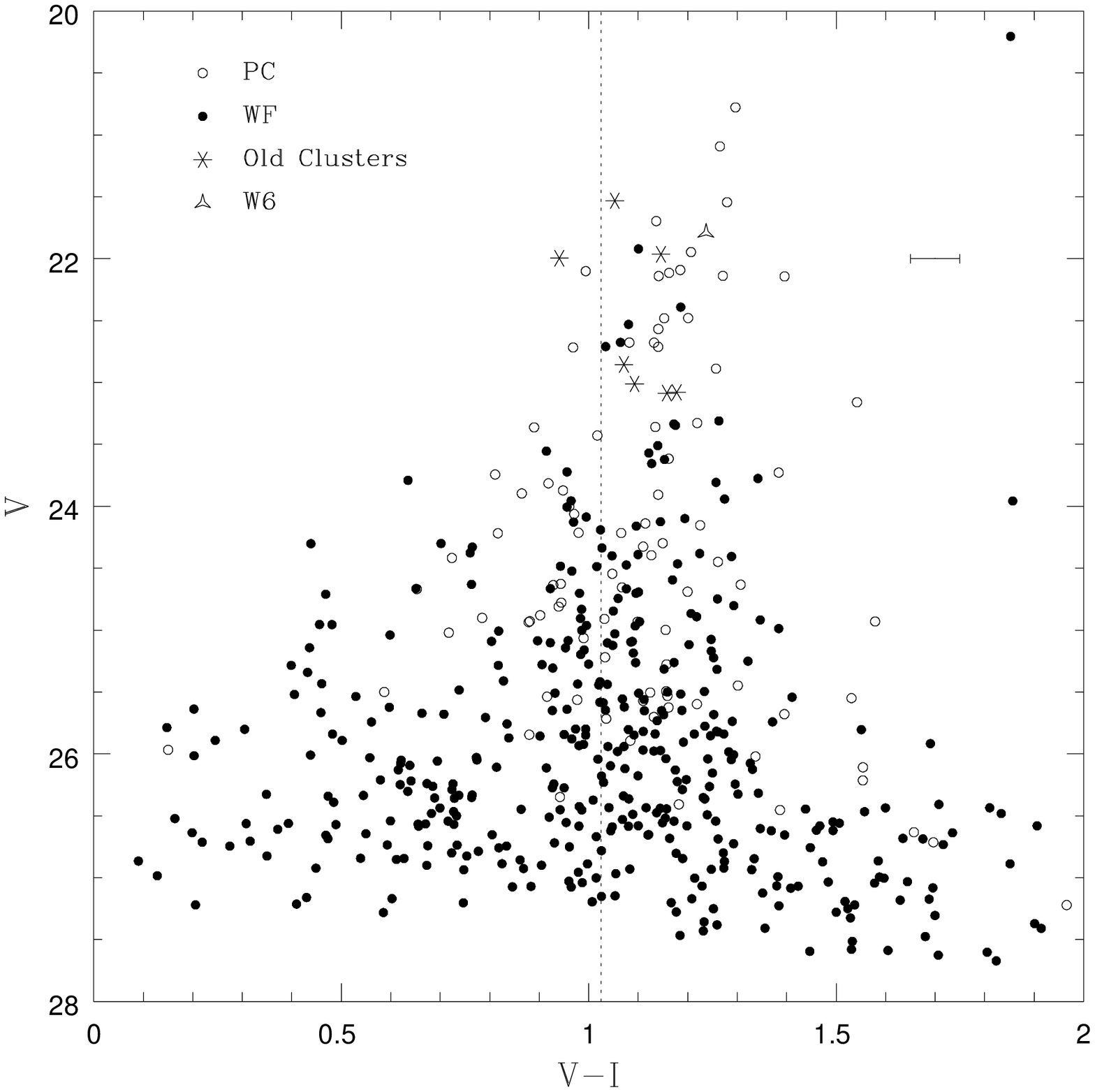}
\figcaption[strader.fig6.ps]{\label{fig:cmdf2}A color-magnitude diagram of all GCs detected in the HST survey by W02. The open circles are objects detected
on the PC chip, and are too close to the galaxy center for spectroscopy. The filled circles are objects detected on the three WF chips. Stars
mark the positions of the seven old GCs in our sample; a triangle marks the intermediate-age GC W6. The dashed line denotes the photometric color cut
($V-I$ = 1.025) adopted by W02 to separate blue and red clusters. A typical error bar for the clusters in our sample is shown.}

\newpage

\epsfxsize=14cm
\epsfbox{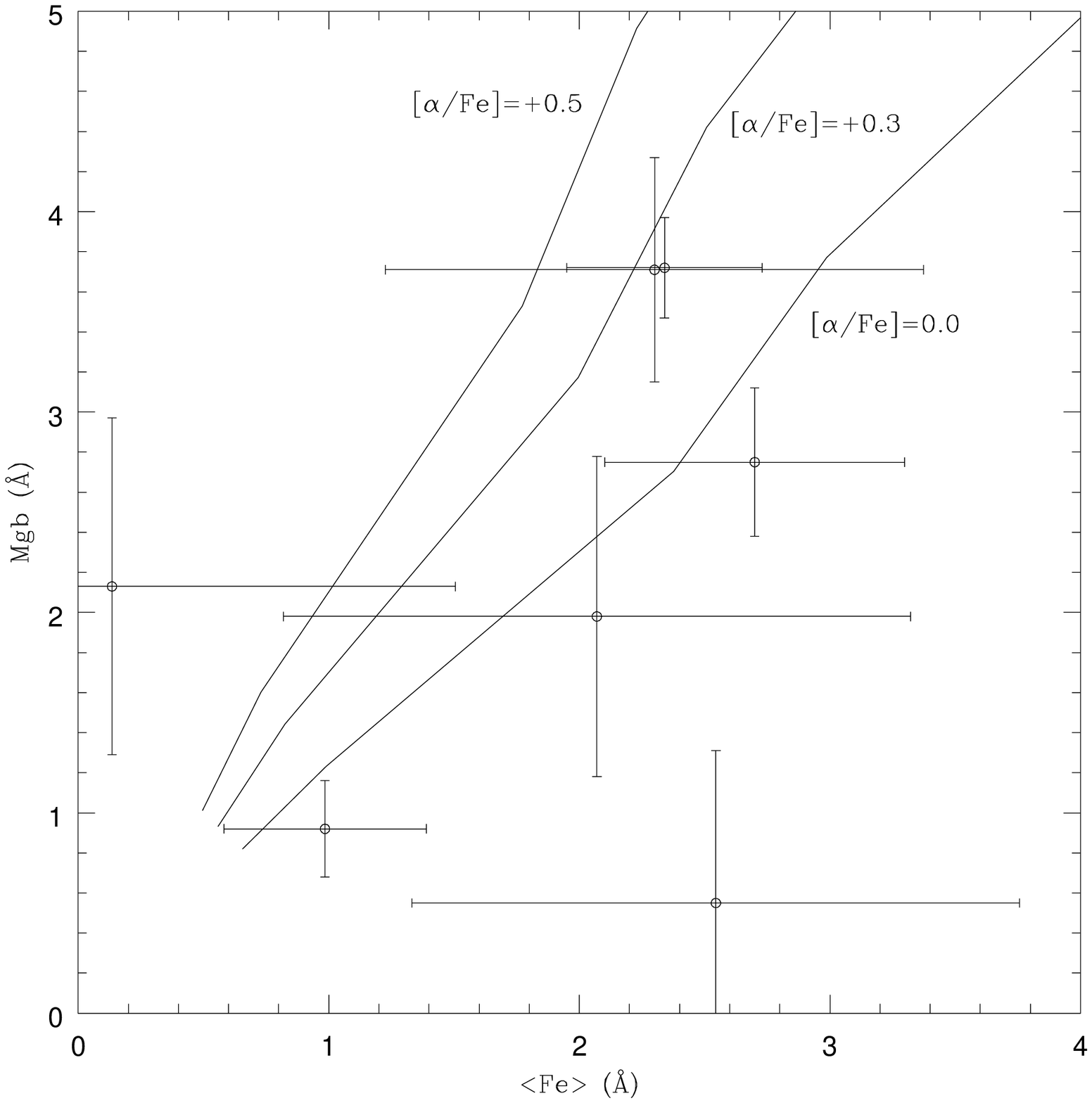}
\figcaption[strader.fig7.ps]{\label{fig:alla}Comparison of Mgb and $<$Fe$>$ for the eight old GCs. 14 Gyr isochrones from Thomas et al.,
varying in [$\alpha$/Fe], are superimposed.}

\epsfxsize=14cm
\epsfbox{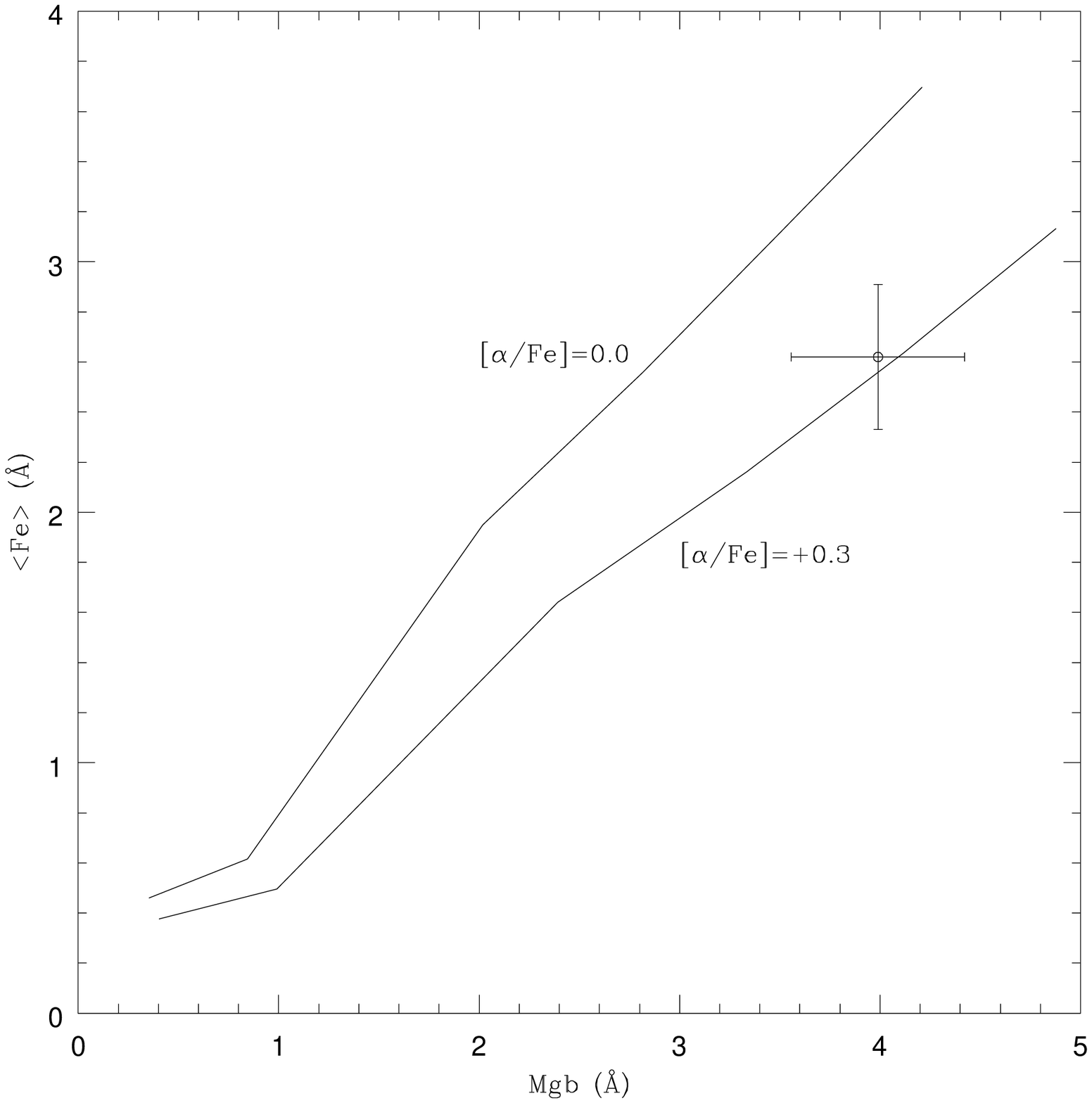}
\figcaption[strader.fig8.ps]{\label{fig:w6a}Comparison of Mgb and $<$Fe$>$ for the intermediate-age cluster W6. 3 Gyr isochrones from Thomas et
al., varying in [$\alpha$/Fe], are superimposed.}

\begin{deluxetable}{lcccccccc}
\tablewidth{0 pt}
\rotate
\tablecaption{ \label{tab:ident}
  Basic Data for Globular-Cluster Candidates in NGC 3610}
\tablehead{ID \tablenotemark{a} & R.A. \tablenotemark{a} & Dec. \tablenotemark{a} & $V$ \tablenotemark{a} & $V - I$ \tablenotemark{a} &
RV \tablenotemark{b} & Proj. Radius \tablenotemark{a,c} & Exp. Time \tablenotemark{d}  & S/N \tablenotemark{e} \\
	&	(hr:min:sec)	&	($^{\circ}$:\arcmin:\arcsec)	&	(mag)	&	(mag)	& (km s$^{-1}$)	& (\arcsec) & (min) & }
\startdata

W3  & 11:18:27.96 & 58:47:16.05 & $21.53\pm0.05$ & $1.05\pm0.05$ & $1831\pm20$  & 22.0 & 300 & 24.7 \\
W6  & 11:18:28.03 & 58:47:05.11 & $21.79\pm0.05$ & $1.24\pm0.05$ & $1799\pm17$  & 22.7 & 560 & 30.4 \\
W9  & 11:18:33.84 & 58:46:53.75 & $21.96\pm0.05$ & $1.15\pm0.05$ & $1825\pm22$ & 69.3 & 560 & 36.7 \\
W10 & 11:18:32.95 & 58:46:15.41 & $22.00\pm0.05$ & $0.94\pm0.05$ & $1774\pm32$ & 82.0 & 560 & 36.5 \\
W26* & 11:18:33.52 & 58:47:42.22 & $23.58\pm0.03$ & $1.22\pm0.04$ & \nodata	 & 72.4 & 260 & 8.5  \\
W28 & 11:18:24.71 & 58:45:53.33 & $22.85\pm0.05$ & $1.07\pm0.05$ & $1712\pm32$ & 77.9 & 260 & 10.4 \\
W30 & 11:18:30.02 & 58:47:28.42 & $23.01\pm0.05$ & $1.09\pm0.05$ & $1724\pm31$ & 41.2 & 560 & 13.4 \\
W31 & 11:18:26.45 & 58:46:29.29 & $23.08\pm0.05$ & $1.18\pm0.05$ & $1755\pm21$ & 42.9 & 560 & 12.3 \\
W32 & 11:18:32.75 & 58:46:03.85 & $23.09\pm0.05$ & $1.16\pm0.05$ & $1732\pm24$ & 89.2 & 560 & 12.9 \\

\enddata
\tablenotetext{a}{Taken from Whitmore et al.\ (2002), except for W26*, which is from Whitmore et al.\ (1997).}
\tablenotetext{b}{Heliocentric radial velocities determined in this work.}
\tablenotetext{c}{At the distance of NGC 3610, a projected radius of 10\arcsec\ corresponds to 1.46 kpc.}
\tablenotetext{d}{Total exposure time over both observing runs.}
\tablenotetext{e}{Average signal-to-noise ratio per resolution element over the wavelength range 4700--5300 \AA.}

\end{deluxetable} 

\begin{deluxetable}{lrrrrrrrrrr}
\tablewidth{0pt}
\rotate
\tabletypesize{\footnotesize}
\tablecaption{[Fe/H] Estimates from Individual BH90 Indices
	\label{tab:metal}}
\tablehead{ ID & $\Delta$ & $\textrm{Mg}_{2}$ & MgH & G Band & CNB & Fe5270 & CNR & H + K}
\startdata

W3 & $-1.05\pm0.37$ & $-0.71\pm0.35$ & $-0.18\pm0.50$ & $-1.01\pm0.40$ & $-0.39\pm0.37$ & $-0.35\pm0.64$ & $-0.96\pm0.47$ & $-1.12\pm0.51$ \\
W6 & $-1.16\pm0.37$ & $0.00\pm0.35$ & $0.60\pm0.49$ & $-0.41\pm0.35$ & $-0.32\pm0.35$ & $-0.32\pm0.63$ & $-0.42\pm0.46$ & $-.64\pm0.45$ \\   
W9 & $-1.55\pm0.37$ & $-0.66\pm0.34$ & $-0.65\pm0.49$ & $-1.33\pm0.35$ & $-1.71\pm0.36$ & $-0.57\pm0.62$ & $-1.12\pm0.46$ & $-1.81\pm0.45$ \\ 
W10 & $ -1.39\pm0.37$ & $-1.65\pm0.34$ & $-1.39\pm0.49$ & $-1.28\pm0.34$ & $-1.99\pm0.34$ & $-1.52\pm0.63$ & $-1.26\pm0.46$ & $-2.51\pm0.44$   \\
W28 & \nodata & $-1.75\pm0.40$ & $-0.43\pm0.62$ & $-0.22\pm0.61$ & \nodata & $-0.23\pm0.78$ & $-2.57\pm0.50$ & $0.23\pm0.76$ \\
W30 & $-2.00\pm0.37$ & $-0.53\pm0.39$ & $0.53\pm0.58$ & $0.01\pm0.46$ & $3.66\pm0.47$ & $-0.09\pm0.73$ & $-0.52\pm0.50$ & $1.29\pm0.47$ \\
W31 & $-2.11\pm0.37$ & $-1.24\pm0.38$ & $-1.38\pm0.56$ & $0.67\pm0.48$ & $0.98\pm0.52$ & $-0.49\pm0.77$ & $-1.02\pm0.49$ & $-1.85\pm0.51$ \\
W32 & $-1.82\pm0.37$ & $-1.28\pm0.39$ & $-1.11\pm0.57$ & $-1.22\pm0.56$ & $-0.41\pm0.44$ & $0.12\pm0.72$ & $-1.30\pm0.49$ & $-0.12\pm0.50$ \\

\enddata
\end{deluxetable}

\begin{deluxetable}{lrrrrrrrrrr}
\tabletypesize{\footnotesize}
\tablecaption{Spectroscopic \& Photometric [Fe/H] Values                    
        \label{tab:metal2}}
\tablehead{ ID & [Fe/H] (spec.) & [Fe/H] (phot.)}
\startdata

W3 & $-0.71\pm0.15$ & $-1.07\pm0.48$ \\
W6 & $-0.29\pm0.19$ & $-0.45\pm0.52$ \\
W9 & $-1.18\pm0.20$ & $-0.74\pm0.50$ \\
W10 & $-1.65\pm0.18$ & $-1.43\pm0.46$ \\
W28 & $-1.02\pm0.57$ & $-1.00\pm0.49$ \\
W30 & $0.40\pm0.69$ & $-0.94\pm0.51$ \\
W31 & $-0.77\pm0.48$ & $-0.64\pm0.50$ \\
W32 & $-0.89\pm0.27$ & $-0.71\pm0.50$ \\

\enddata
\end{deluxetable}

\begin{deluxetable}{lcccccccccc}
\tablewidth{0pt}
\rotate
\tabletypesize{\footnotesize}
\tablecaption{Lick/IDS Indices\tablenotemark{a}
	\label{tab:indic}}
\tablehead{ID &  H$\beta$ & H$\gamma_{A}$ & H$\delta_{A}$ & Ca4227 & G4300 & Fe5270 & Fe5335 & $\rm{Mg}_{2}$ & Mgb   & CN2 \\
	      &	 (\AA)	  &    (\AA)      & (\AA)         & (\AA)  & (\AA) & (\AA)  & (\AA)  & (mag)             & (\AA) & (mag)}

\startdata

W3 & $1.17\pm0.32$ & $-3.16\pm0.55$ & $-0.25\pm0.73$ & $-0.23\pm0.38$ & $3.85\pm0.60$ & $2.53\pm0.37$ & $2.87\pm0.47$ & $0.16\pm0.01$ & $2.75\pm0.37$ & $0.06\pm0.02$ \\
W6 & $2.59\pm0.27$ & $-2.90\pm0.33$ & $-1.24\pm0.51$ & $1.14\pm0.27$ & $5.16\pm0.36$ & $2.66\pm0.28$ & $2.58\pm0.33$ & $0.23\pm0.01$ & $3.99\pm0.29$ & $0.13\pm0.01$ \\ 
W9 & $1.42\pm0.21$ & $-2.22\pm0.39$ & $1.33\pm0.38$ & $0.82\pm0.22$ & $2.89\pm0.39$ & $2.46\pm0.25$ & $2.22\pm0.30$ & $0.16\pm0.01$ & $3.72\pm0.25$ & $0.00\pm0.01$ \\
W10 & $2.34\pm0.22$ & $-0.90\pm0.34$ & $1.09\pm0.41$ & $0.24\pm0.19$ & $3.05\pm0.34$ & $0.82\pm0.27$ & $1.15\pm0.30$ & $0.06\pm0.01$ & $0.92\pm0.24$ & $0.01\pm0.01$ \\
W28 & $2.28\pm0.81$ & $-0.89\pm1.25$ & $6.16\pm0.97$ & $2.61\pm0.70$ & $5.53\pm1.21$ & $3.07\pm0.79$ & $-2.80\pm1.12$ & $0.06\pm0.02$ & $2.13\pm0.84$ & $-0.05\pm0.04$ \\
W30 & $1.50\pm0.62$ & $-3.99\pm1.06$ & $0.56\pm0.94$ & $0.47\pm0.27$ & $5.43\pm0.75$ & $2.57\pm0.67$ & $2.03\pm0.84$ & $0.18\pm0.02$ & $3.71\pm0.56$ & $0.11\pm0.03$ \\
W31 & $0.92\pm0.63$ & $-3.82\pm0.90$ & $5.82\pm1.09$ & $-0.18\pm0.73$ & $5.95\pm0.85$ & $2.52\pm0.79$ & $1.62\pm0.97$ & $0.11\pm0.02$ & $1.98\pm0.80$ & $0.01\pm0.02$ \\
W32 & $0.93\pm0.62$ & $-3.14\pm1.04$ & $-2.14\pm0.61$ & $0.55\pm0.55$ & $2.29\pm1.13$ & $3.22\pm0.64$ & $1.87\pm1.03$ & $0.10\pm0.02$ & $0.55\pm0.76$ & $-0.01\pm0.03$ \\

\enddata

\tablenotetext{a}{These indices were measured according to the definitions in Trager et al. (1998) and Worthey \& Ottaviani (1997).}

\end{deluxetable}

\end{document}